\documentstyle[12pt]{article}
\newcommand{\beq}{\begin{equation}}
\newcommand{\eeq}{\end{equation}}
\newcommand{\beqa}{\begin{eqnarray}}
\newcommand{\eeqa}{\end{eqnarray}}
\newcommand{\ba}{\begin{array}}
\newcommand{\ea}{\end{array}}

\begin{document}

\begin{center}
{\Large \bf Different Facets of Chaos \\in Quantum Mechanics}
\vskip 0.5 truecm
{\bf V.R. Manfredi}$^{(1)}$ and {\bf L. Salasnich}$^{(2)}$ 
\vskip 0.5 truecm
$^{(1)}$Dipartimento di Fisica ``G. Galilei'', Universit\`a di Padova, \\
Istituto Nazionale di Fisica Nucleare, Sezione di Padova, \\
Via Marzolo 8, I-35131 Padova, Italy 
\vskip 0.3 truecm
$^{(2)}$Istituto Nazionale per la Fisica della Materia, Unit\`a di Milano, \\
Dipartimento di Fisica, Universit\`a di Milano, \\
Via Celoria 16, I-20133 Milano, Italy 
\end{center}

\vskip 1. truecm

\begin{center}
{\bf Abstract}
\end{center}
\vskip 0.5 truecm 
\par  
Nowadays there is no universally accepted definition of quantum chaos. 
In this paper we review and critically 
discuss different approaches to the subject, 
such as Quantum Chaology and the Random Matrix Theory. Then we analyze 
the problem of dynamical chaos and the time scales associated 
with chaos suppression in quantum mechanics. 

\vskip 0.5 truecm 
PACS Numbers: 05.45.+b, 03.65.Bz 

\newpage

\section{Introduction}

The aim of this paper is to review and discuss various 
definitions and approaches to {\it quantum chaos}.$^{1-5}$ 
\par 
As yet, there is no universally accepted 
definition of quantum chaos. On the contrary, the meaning of classical 
chaos is beyond question. In classical mechanics a trajectory 
${\bf z}(t)$ in the phase space 
$\Omega$ is chaotic if its maximal Lyapunov exponent $\lambda$ is positive. 
The Lyapunov exponent is defined as  
\beq 
\lambda =\lim_{t\to \infty}{1\over t}\ln{|{\bf \omega} (t)|} \; ,
\eeq 
where ${\bf \omega} (t)$ is a tangent vector to ${\bf z}(t)$ with the
condition that $|{\bf \omega} (0)|=1$. 
The exponential instability of chaotic trajectories 
implies a continuous frequency spectrum of motion. 
The continuous spectrum, in turn, implies correlation decay; this property, 
which is called {\it mixing} in ergodic theory, is the most important 
property of dynamical motion for the validity of the statistical 
description.$^{6-8}$ 
\par 
The problem of quantum chaos arose because 
the above mentioned condition of continuous spectrum for classical chaos is 
violated in quantum mechanics. Indeed the energy and the frequency 
spectrum of any quantum motion, bounded in phase space, are 
always discrete due to the non-commutative geometry 
(discreteness) of the phase space. According 
to the theory of dynamical systems, such motion corresponds 
to the limiting case of regular motion. It means that there is no 
classical-like chaos at all in quantum mechanics.$^{1-5}$ 
\par 
Nevertheless, we shall show that it is reasonable and useful 
to apply the word {\it chaos} also in quantum mechanics. 
In the first part of this article some definitions of quantum chaos 
for stationary systems are discussed, while in the 
second part, the time evolution of classical and quantum systems is  
compared. 

\section{Quantum Chaology and Spectral Statistics} 

As is well known, in the study of the transition from order to chaos 
in classical systems a useful tool is the examination 
of the phase space properties, such as the Poincar\`e sections.$^{6}$ 
Such plots are {\it not} directly available in the case of quantum systems. 
In many papers, the Berry definition of quantum chaos is adopted: 
"Quantum Chaology is the study of semiclassical, but not classical, 
behaviour characteristic 
of systems whose classical motion exhibits chaos".$^{8}$ 
\par 
The idea, also suggested by authors like Percival 
and Gutzwiller,$^{2}$ 
is to connect the behaviour of the eigenvalues and eigenfunctions 
of a quantum system to the different structure of the phase space 
of the corresponding classical system in the regular and chaotic region. 
\par 
In the context of quantum chaology, the spectral 
statistics of the energy levels are of great importance. 
Mehta defined: "A spectral statistic is a quantity which can be calculated 
from an observed sequence of levels alone, without other information 
and whose average value and variance are known from the theoretical 
model. A suitable statistic is one which is sensitive for the properties 
to be compared or distinguished and is insensitive for 
other details".$^{9}$  In particular, 
it has been found that the spectral statistics
of systems with underlying 
classical chaotic behaviour and time-reversal symmetry agree with 
the predictions of the Gaussian Orthogonal Ensemble (GOE) of 
Random Matrix Theory (RMT),$^{9}$ whereas quantum analogs of classically 
integrable systems display the characteristics 
of the Poisson statistics.$^{1-5}$ 
Note that if the chaotic system is without time-reversal symmetry then, 
instead of the GOE, it follows the predictions of the Gaussian 
Unitary Ensemble (GUE). 
\par 
The most used spectral statistics of the energy levels 
are $P(s)$ and $\Delta_3(L)$. $P(s)$ is 
the distribution of nearest-neighbour spacings 
$s_i=({\tilde E}_{i+1}-{\tilde E}_i)$ 
of the unfolded levels ${\tilde E}_i$. 
It is obtained by accumulating the number of spacings that lie within 
the bin $(s,s+\Delta s)$ and then normalizing $P(s)$ to unit. 
As shown by Berry,$^{10,11}$ 
for quantum systems whose classical analogs are integrable, 
$P(s)$ is expected to follow the Poisson distribution 
\beq 
P(s)=\exp{(-s)} \; . 
\eeq 
On the other hand, quantal analogs of chaotic systems 
exhibit the spectral properties of GOE with 
\beq 
P(s)= {\pi \over 2} s \exp{(-{\pi \over 4}s^2)} \; , 
\eeq 
which is the so-called Wigner distribution. Note that 
for systems without time-reversal symmetry the GUE predicts 
$P(s)=(32/\pi^2)s^2\exp{(-4s^2/\pi)}$. 
\par
The statistic $\Delta_3(L)$ is defined, for a fixed interval $(-L/2,L/2)$, as
the least-square deviation of the staircase function $N(E)$ from
the best straight line fitting it:
$$
\Delta_{3}(L)={1\over L}\min_{A,B}\int_{-L/2}^{L/2}[N(E)-AE-B]^2 dE \; ,
$$
where $N(E)$ is the number of levels between E and zero for positive
energy, between $-E$ and zero for negative energy. The $\Delta_{3}(L)$ 
statistic provides a measure of the degree of rigidity of the
spectrum: for a given interval L, the smaller $\Delta_{3}(L)$ is,
the stronger is the rigidity, signifying the long-range 
correlations between levels. For this statistic 
the Poissonian prediction is 
\beq 
\Delta_3(L)= {L\over 15} \; . 
\eeq
The GOE predicts the same behaviour for $L<<1$; instead 
for $L>>1$ it gives 
\beq 
\Delta_{3}(L)={1\over \pi^2}\log{L} \; . 
\eeq 
In the GUE case one has $\Delta_{3}(L)=(1/2\pi^2)\log{L}$. 
It is useful to remember that Berry$^{12}$ has shown that $\Delta_3(L)$ 
deviates from the universal predictions of RMT for very large $L$. 
\par 
Another probe, which is generally regarded (see for instance Ref. 14) 
as very sensitive to the structure of chaotic states, is the 
transition probability. For reasons of space, we mention only the 
pioneering work of French and his coworkers$^{13}$ and two more recent 
works concerning the interacting-boson model$^{14}$ (IBM) and the 
three-level Lipkin, Meshkov, Glick (LMG) model.$^{15}$ 
In both works the results 
based on the transition probabilities between eigenstates of the 
system completely agree with the spectral statistics $P(s)$ and $\Delta_3(L)$. 
\par
It is important to stress that even though the classical system 
is not known, to distinguish between ordered and 
chaotic states, the spectral statistics and the 
transition probabilities can be used.$^{16-19}$ 

\section{From Poisson to GOE Transition: Comparison 
with Experimental Data} 

\par
The agreement between the classical order-chaos transition and the 
quantal Poisson-GOE transition has been tested in many 
theoretical models, ranging from simple billiards$^{20-22}$ to 
more realistic systems like nuclei$^{23-26}$ and 
elementary particles.$^{27-29}$
\par 
In this section we shall compare the transition Poisson-GOE with 
the experimental data of two different systems: the atomic nuclei 
and the Hydrogen atom in a static magnetic field. 

\subsection{Atomic Nuclei}

In atomic nuclei, as in other many-body systems, ordered 
and chaotic states generally coexist.$^{17}$

\vskip 0.5 truecm 
\par
{\bf a) The Low Energy Region} 
\vskip 0.5 truecm 
\par
The behaviour of spectral statistics near the ground state has been 
studied by Garret, German, Courtney and Espino$^{17}$ and Shriner, 
Mitchell and Von Egidy$^{17}$. The main results of these 
authors have been shown in Figure 1 and Figure 2. 
As can be seen from the figures, in the low energy region 
the spectral statistics are in agreement with the Poisson 
ensemble or intermediate between Poisson and GOE. 

\vskip 0.5 truecm 

\par
{\bf b) The High Energy Region}
\vskip 0.5 truecm 
\par 
Neutron resonance spectroscopy on a heavy even-even nucleus
typically leads to the identification of about 150 to 170 s-wave
resonances with $J^{\pi}={1\over 2}^{+}$ located 8-10
MeV above the ground state of the compound system, with average
spacings around 10 eV and average total widths around 1 eV. Proton
resonance spectroscopy yields somewhat shorter sequences of levels
with fixed spin and parity, with typically 60 to 80 members. 
\par 
For the statistical analysis, it is essential that the sequences be pure
(no admixture of levels with different spin or parity) and
complete (no missing levels). Only such sequences were considered
by Haq, Pandey and Bohigas.$^{30}$ Scaling each sequence to the same 
average level spacing and lumping together all sequences one 
leads to the ''Nuclear
Data Ensemble" (NDE), which contains 1726 level spacings. 
\par 
As shown in Figure 3, the agreement between the experimental data and the 
GOE predictions is surprisingly good (in the GOE model there are no 
free parameters).
 
\subsection{The Hydrogen Atom in the Strong Magnetic Field}

We now discuss the local statistical properties 
of energy levels of a Hydrogen atom in a uniform strong magnetic 
field. The Hamiltonian of the system is given by 
\beq
H={p^2\over 2m}-{e^2\over r}-{qB\over 2m}L_z 
+{q^2 B^2\over 8 m} (x^2+y^2) \; ,
\eeq
where the magnetic field $B$ breaks the time-reversal symmetry. 
Although Eq. (6) is {\it not} time-reversal invariant, 
it can easily be written in a time-reversal invariant form.$^{31}$ 
In fact, the paramagnetic interaction ${qB\over 2m}L_z$ simply 
shifts the whole series of levels with a fixed quantum number $M$ 
(eignenstate of $L_z$ with eigenvalue $M\hbar$) and 
can be taken into account in the standard way: the Zeeman effect. 
\par
The Hamiltonian (6) written 
in atomic units $m=|q|=4\pi \epsilon_0 =\hbar =1$ is 
\beq
H={p^2\over 2}-{1\over r}+{\gamma^2 \over 8} (x^2+y^2) \; ,
\eeq
where $\gamma=B/B_c$ is the magnetic field in atomic units and 
$B_c=2.35 \cdot 10^5$ Tesla. 
This equation can be numerically solved for different 
values of the scaled energy $\epsilon = E/(2\gamma)^{2/3}$. 
Once the eigenvalues have been obtained, the spectral 
statistics $P(s)$ and $\Delta_3(L)$ can be calculated. 
Figure 4 shows the function $P(s)$ for different values 
of the scaled energy $\epsilon$. Increasing $\epsilon$, 
a smooth Poisson-GOE transition can be observed. 
Figure 5 shows the spectral rigidity $\Delta_3(L)$ in three 
different energy intervals. For this statistic the transition 
Poisson-GOE is also very clear.  
\par 
In addition, a comparison has been made between the theoretical 
energy levels and the experimental ones; the agreement 
is excellent. 

\section{Quantum Chaos and Field Theory} 

\par
In the last few years there has been much interest in chaos 
in field theories. It is now well known that the spatially uniform limits 
of scalar electrodynamics and Yang-Mills theory exhibit classical 
chaotic motion.$^{34}$ 
In this section we discuss quantum chaos in a field-theory schematic model, 
namely the spatially homogeneous SU(2) Yang-Mills-Higgs 
(YMH) system.$^{27-29}$ 
The Lagrangian density of the SU(2) YMH system is given by 
\beq
L={1\over 2}(D_{\mu}\phi )^+(D^{\mu}\phi ) -V(\phi ) 
-{1\over 4}F_{\mu \nu}^{a}F^{\mu \nu a} \; ,
\eeq
where
\beq
(D_{\mu}\phi )=\partial_{\mu}\phi - i g A_{\mu}^b T^b\phi 
\; ,
\eeq
\beq
F_{\mu \nu}^{a}=\partial_{\mu}A_{\nu}^{a}-\partial_{\nu}A_{\mu}^{a}+
g\epsilon^{abc}A_{\mu}^{b}A_{\nu}^{c} \; ,
\eeq
with $T^b=\sigma^b/2$, $b=1,2,3$, generators of the SU(2) algebra, 
and where the potential of the scalar field (the Higgs field) is
\beq
V(\phi )=\mu^2 |\phi|^2 + \lambda |\phi|^4 \; .
\eeq 
In the (2+1)-dimensional Minkowski space ($\mu =0,1,2$) and with 
spatially homogeneous Yang-Mills and the Higgs fields
\beq
\partial_i A^a_{\mu} = \partial_i \phi = 0 \; , \;\;\;\; i=1,2
\eeq
one considers the system in the region in which space fluctuations of 
fields are negligible compared to their time fluctuations. 
\par
In the gauge $A^a_0=0$ and using the real triplet representation for the 
Higgs field one obtains 
$$
L={\dot{\vec \phi}}^2 +
{1\over 2}({\dot {\vec A}}_1^2+{\dot {\vec A}}_2^2) 
-g^2 [{1\over 2}{\vec A}_1^2 {\vec A}_2^2 
-{1\over 2} ({\vec A}_1 \cdot {\vec A}_2)^2+
$$
\beq
+({\vec A}_1^2+{\vec A}_2^2){\vec \phi}^2 
-({\vec A}_1\cdot {\vec \phi})^2 -({\vec A}_2 \cdot {\vec \phi})^2 ]
-V( {\vec \phi} ) \; ,
\eeq
where ${\vec \phi}=(\phi^1,\phi^2,\phi^3)$, 
${\vec A}_1=(A_1^1,A_1^2,A_1^3)$ and ${\vec A}_2=(A_2^1,A_2^2,A_2^3)$. 
\par
When $\mu^2 >0$, the potential $V$ has a minimum at $|{\vec \phi}|=0$, 
but for $\mu^2 <0$ the minimum is at 
$$
|{\vec \phi}_0|=\sqrt{-\mu^2\over 4\lambda }=v \; ,
$$
which is the non zero Higgs vacuum. This vacuum is degenerate, 
and after spontaneous symmetry breaking the physical vacuum can be 
chosen ${\vec \phi}_0 =(0,0,v)$. If $A_1^1=q_1$, $A_2^2=q_2$ 
and the other components of the Yang-Mills fields are zero, 
in the Higgs vacuum the Hamiltonian of the system reads 
\beq
H={1\over 2}(p_1^2+p_2^2)
+g^2v^2(q_1^2+q_2^2)+{1\over 2}g^2 q_1^2 q_2^2 \; ,
\eeq
where $p_1={\dot q_1}$ and $p_2={\dot q_2}$. Here $w^2=2 g^2v^2$ is the 
mass term of the Yang-Mills fields. This YMH Hamiltonian is 
a toy model for classical non-linear dynamics, with the attractive feature 
that the model emerges from particle physics. 
At low energy the motion near the minimum of the potential 
\beq 
V(q_1,q_2)=g^2 v^2 (q_1^2+q_2^2)+{1\over 2} g^2 q_1^2 q_2^2 \; ,
\eeq 
where the Gaussian curvature is positive, is periodic or quasiperiodic and is 
separated from the instability region by a line of zero curvature; 
if the energy is increased, the system will be for some initial conditions 
in a region of negative curvature, where the motion is chaotic. 
According to this scenario, the energy $E_c$ of chaos-order transition 
is equal to the minimum value of the line of zero Gaussian 
curvature $K(q_1 ,q_2 )$ on the potential-energy surface. 
It is easy to show that the minimal energy on the 
zero-curvature line is given by:
\beq
E_c=V_{min}(K=0,\bar{q_1})=6 g^2 v^4 \; , 
\eeq
and by inverting this equation one obtains $v_c=(E /6g^2)^{1/4}$. 
There is an order-chaos transition by increasing the energy $E$ 
of the system and a chaos-order transition by increasing 
the value $v$ of the Higgs field in the vacuum. Thus, 
there is only one transition regulated by the sole parameter $E/(g^2v^4)$. 
\par
It is important to point out that 
{\it in general} the curvature 
criterion guarantees only a {\it local instability}$^{16)}$ 
and should therefore be combined with the Poincar\`e sections$^{17)}$. 
Chaotic regions on the surface of the section 
are characterized by a set of randomly distributed points, 
and regular regions by dotted or solid curves. 
Figure 6 shows the Poincar\`e sections, which 
confirm the analytical predictions of the curvature criterion: 
the critical value of the onset of chaos is 
in very good agreement with the Poincar\`e sections. 
\par
In quantum mechanics the generalized coordinates of the YMH system 
satisfy the usual commutation rules $[{\hat q}_k,{\hat p}_l]=i\delta_{kl}$, 
with $k,l=1,2$. Introducing the creation and destruction operators
\beq
{\hat a}_k=\sqrt{\omega \over 2}{\hat q}_k + 
i \sqrt{1\over 2\omega}{\hat p}_k \; ,
\;\;\;\;
{\hat a}_k^+ = \sqrt{\omega \over 2}{\hat q}_k - 
i \sqrt{1\over 2\omega}{\hat p}_k \; ,
\eeq
the quantum YMH Hamiltonian can be written
\beq
{\hat H}={\hat H}_0 + {1\over 2} g^2 {\hat V} \; ,
\eeq
where
\beq
{\hat H}_0= \omega ({\hat a}_1^+ {\hat a}_1 + {\hat a}_2^+ {\hat a}_2 + 1) \; ,
\eeq
\beq
{\hat V}= {1 \over 4 \omega^2} ({\hat a}_1 +{\hat a}_1^+)^2 
({\hat a}_2 +{\hat a}_2^+)^2 \; ,
\eeq
with $\omega^2 = 2 g^2 v^2$ and $[{\hat a}_k,{\hat a}_l^+] = \delta_{kl}$, 
$k,l=1,2$. 
If $|n_1 n_2>$ is the basis of the occupation numbers of the two 
harmonic oscillators, the matrix elements are
\beq
<n_{1}^{'}n_{2}^{'}|{\hat H}_0|n_{1}n_{2}>= \omega (n_1+n_2+1) 
\delta_{n_{1}^{'}n_{1}} \delta_{n_{2}^{'}n_{2}} \; ,
\eeq
and
$$
<n_{1}^{'}n_{2}^{'}|{\hat V}|n_{1}n_{2}>=
{1 \over 4 \omega^2}
[\sqrt{n_{1}(n_{1}-1)} \delta_{n^{'}_{1}n_{1}-2}
+\sqrt{(n_{1}+1)(n_{1}+2)}\delta_{n^{'}_{1}n_{1}+2}+
(2n_{1}+1)\delta_{n^{'}_{1}n_{1}}]\times 
$$
\beq
\times[\sqrt{n_2 (n_2-1)}\delta_{n^{'}_2 n_2-2}+ \sqrt{(n_2+1)(n_2+2)}
\delta_{n^{'}_2 n_2+2}+ (2n_2+1)\delta_{n^{'}_2 n_2}] \; .
\eeq 
Figure 7 shows the $P(s)$ distribution 
for different values of the parameter $v$. 
The figure shows a Wigner-Poisson transition by increasing the value $v$ 
of the Higgs field in the vacuum. The $P(s)$ distribution is 
fitted by the Brody function 
\beq
P(s,\omega)=\alpha (\omega +1) s^{\omega} \exp{(-\alpha s^{\omega+1})} \; ,
\eeq
with 
\beq
\alpha = \big( \Gamma [{\omega +2\over \omega+1}] \big)^{\omega +1} \; .
\eeq 
This function interpolates between the Poisson distribution ($\omega =0$) 
of integrable systems and the Wigner distribution ($\omega =1$) of 
chaotic ones, and thus the parameter $\omega$ can be used as a simple 
quantitative measure of the degree of chaoticity. 
By using the P(s) distribution and the Brody function 
it is possible to give a quantitative measure 
of the degree of quantal chaoticity of the system. 
The numerical calculations of Figure 6 and 7 clearly show the quantum 
chaos-order transition and its connection to the classical one.  

\section{Alternative Approaches to Quantum Chaos}

A different approach to quantum chaos has been discussed by Sakata and 
his coworkers.$^{35}$ The example discussed by Sakata 
is a system of an even number of 
fermions transformed into a boson system by means of the boson 
expansion theory,$^{36}$ where the boson system is described by 
K-kinds of boson operators ($B_j$, $B_j^+$; j=1,...K). 
\par
The main idea of Sakata$^{35}$ is that, just as in the classical theory 
a dissolution of integrability (with the KAM mechanism)  
simply means the onset of chaotic motion,$^{7,8}$ 
in quantum systems a dissolution of quantum numbers may indicate the 
onset of quantum chaos. In accordance with the above definition 
of quantum chaos, we may classify the eigenstates $|i>$ 
of the many-body Hamiltonian into 
three characteristic cases with the aid of the $(\mu ,\nu)$-basis states, 
defined as $|\mu , \nu> =|\mu_1 ... \mu_{L} , \nu_{L+1} ... \nu_K >$, 
which are specified by $K$-kinds of quantum numbers.  
\\
1) \underbar{Quantum integrable states}: 
If one finds one of the $|\mu , \nu>$-basis states for a given eigenstate 
$|i>$ satisfying $|<\mu , \nu | i>|^2=1$, 
then $|i>$ is classified as a {\it quantum 
integrable state}, see Figure 8(a). 
\\
2) \underbar{Quantum KAM states}: 
If $|i>$ is described perturbatively starting from the 
$|\mu , \nu >$-basis state, then it is a 
{\it quantum KAM state}, see Figure 8(b). 
\\
3) \underbar{Quantum chaotic states}: 
If $|i>$ is {\it not} described perturbatively starting from 
the $|\mu , \nu>$-basis state, then it is regarded as a 
{\it quantum chaotic state}, see Figure 8(c). 
\par
For further discussion of Sakata's approach see Ref. 35 and references 
quoted therein. 
\par
Recently, another approach to the order-chaos transition 
has been proposed by Soloviev in the framework of nuclear 
structure.$^{37,38}$ Soloviev's main idea 
was to discuss the order to chaos transition 
in terms of the properties of nuclear wave-functions 
and to analyze how the structure of nuclear states changes with 
increasing excitation energy. 
He focused his attention 
on non-rotational states of rigid nuclei. 
The main conclusions are the following: 
\\
1) Order is governed by the large components of the wave function 
of the excited states. 
\\
2) Chaos takes place in the small components of the wave function 
of the nuclear excited states. The excited states are chaotic 
if their wave functions are composed of only small components of 
many-quasiparticle or many-phonon configurations. 
\par
In our opinion these two approaches are quite similar and 
the Sakata approach also gives a simpler picture of the three cases discussed 
(regular, KAM, chaotic). 

\section{Dynamical Quantum Chaos and Time Scales}

In this section we analyze in more detail the previously discussed 
problems of the chaotic time evolution for a quantum system. 
\par 
A quantum system evolves according 
to a linear equation and this is an important feature which makes it 
different from a classical system, for which the equations of motion can be 
nonlinear. On the other hand, the Liouville equation of the density function 
is linear both in classical and quantum mechanics but the evolution 
operator of Liouville has different spectral properties. 
The classical Liouville operator 
has a continuous spectrum and this implies and allows chaotic motion. 
Instead, for bound systems, the quantum Liouville operator 
has a purely discrete spectrum, therefore no long-term 
chaotic behaviour.$^{4,5}$ 
\par 
As shown by Casati, Chirikov and coworkers$^{39,40}$ 
studying toy models like the kicked rotor, 
the time evolution of a quantum state follows 
the classical one, including the chaotic phenomena, 
up to a {\it break time} $t_B$. 
After that, in contrast to classical dynamics, 
we get localization (dynamical localization). 
This means that persistent chaotic behaviour in the evolution 
of the quantum states and observables is not possible. 
Roughly speaking, chaotic behaviour is possible in quantum mechanics only 
as a transient with lifetime $t_B$. The phenomenon of 
localization is clearly illustrated in Figure 9. This plot shows, 
in the case of the so-called standard map, the classical 
(solid curve) and quantum (dotted curve) unperturbed energy as a function of 
$\tau$ (number of map iteractions). 
\par 
The value of the break time $t_B$ depends on the model studied 
and its exact behaviour is still controversial, but 
can be estimated from the Heisenberg indetermination principle as 
\beq 
t_B \simeq {\hbar \over \Delta E} \; ,
\eeq
where $\Delta E$ is the mean spacing of energy levels. 
The discrete spectrum of the Liouville 
operator cannot be resolved if $t<t_B$, i.e. $t_B$ 
is the time at which the quantal evolution (of a wave packet, for example)
"realizes" that the spectrum of the evolution operator is discrete. 
According to the Thomas-Fermi rule, 
$\Delta E \propto \hbar^N$, where N is the number of degrees of freedom, 
i.e. the dimension of the configuration space. 
So, as $\hbar \rightarrow 0$, the break time 
diverges as $t_B \sim \hbar^{1-N}$, and it does so faster, the higher $N$ is. 
\par 
In the case of classical chaos, another time scale, 
{\it the random time scale} $t_R$, much shorter than $t_B$, can be introduced 
to estimate the time at which classical exponential spreading 
reaches the quantal resolution of the phase space. Thus, it 
is the full time for the exponential spreading of the minimum 
initial wave packet. As shown by Berman and Zaslavski$^{41}$ 
the random time (called the breaking time by the two authors) 
follows the logarithmic law 
\beq 
t_R \simeq \lambda^{-1} \ln{\Big( {S\over \hbar} \Big)} \; , 
\eeq 
where $S$ is a classical action and $\lambda$ 
is the Lyapunov exponent of the system. 
In particular, Berman and Zaslavski have studied two different models, 
which allow one to calculate the random time scale $t_R$. 
The first model is a periodically kicked oscillator, which gives 
$t_R=ln{(cost/\hbar )}$. The second is an ensemble of $N$ atoms 
interacting with light in the resonant cavity. 
For this model $t_R = \ln{(cost \; N)}$. 

\subsection{Mean-Field Approximation and Dynamical Chaos}

Let us consider a $N$-body quantum system with 
Hamiltonian ${\hat H}$. The exact time-dependent Schr\"odinger equation 
can be obtained by imposing the quantum last action principle on 
the Dirac action 
\beq 
S= \int dt <\psi (t) | i\hbar 
{\partial \over \partial t} - {\hat H} |\psi (t) > \; ,
\eeq
where $\psi$ is the many-body wavefunction of the system.$^{42}$ Looking 
for stationary points of $S$ with respect to variation of the conjugate 
wavefunction $\psi^*$ gives 
\beq
i\hbar {\partial \over \partial t}\psi = {\hat H}\psi \; .
\eeq 
As is well known, it is usually impossible to obtain the exact solution 
of the many-body Schr\"odinger equation and some approximation must be used. 
In the mean-field approximation the total wavefunction 
is assumed to be composed of independent particles, i.e. it can be 
written as a product of single-particle wavefunctions $\phi_j$. 
In the case of identical fermions, $\psi$ must be antisymmetrized. 
By looking for stationary action with respect to variation of a 
particular single-particle conjugate wavefunction $\phi_j^*$ one finds 
a time-dependent Hartree-Fock equation for each $\phi_j$: 
\beq
i\hbar {\partial \over \partial t}\phi_j = {\delta \over \delta \phi_j^*} 
<\psi | {\hat H}| \psi > = {\hat h} \phi_j \; ,
\eeq
where ${\hat h}$ is a one-body operator.$^{42}$ 
The main point is that, in general, 
the one-body operator ${\hat h}$ is nonlinear. Thus 
the Hartree-Fock equations are non-linear (integro-)differential 
equations. These equations can give rise, in some cases, 
to chaotic behaviour (dynamical chaos) of the mean-field wavefunction. 
\par
In the mean-field approximation the 
mathematical origin of {\it dynamical chaos} resides in the nonlinearity 
of the Hartree-Fock equations. These equations provide an approximate 
description, the best independent-particle description, which 
describes, for a certain time interval, the very complicated 
evolution of the true many-body system. 
Two questions then arise: \\
1) Does this chaotic behavior persist in time? \\
2) What is the best physical situation to observe this kind of nonlinearity? 
\par
To answer the first question, 
it should be stressed that, as shown previously, 
quantum systems evolve according 
to a linear equation. Since the Schr\"odinger equation 
is linear, so is any of its projections. Its time 
evolution follows the classical one, including chaotic
behaviour, up to the break time $t_B$. 
After that, in contrast to the classical dynamics, 
we get dynamical localization. 
This means that persistent chaotic behaviour in the evolution 
of the states and observables is not possible. Nevertheless, 
we have seen that $t_B \sim \hbar^{1-N}$, where $N$ is the number 
of degrees of freedom of the system, thus for a large number 
of particles the break time can be very long. 
\par 
Concerning the second question, it is useful 
to remember that, in the thermodynamic limit, 
i.e. when the number $N$ of particles tends to 
infinity at constant density, the energy spectrum of the system is, 
in general, continuous and true chaotic phenomena are not excluded.$^{43}$  
\par 
When the mean-field theory is a good approximation 
of the exact many-body problem, 
one can use the nonlinear mean-field equations to estimate 
the transient chaotic behaviour of the many-body system. 
An important case where such an approach could be applied is  
the Bose--Einstein condensate of weakly-interacting alkali-metal 
atoms.$^{44}$ In particular, we suggest that the collective oscillations 
of the Bose condensate$^{45,46}$ can give rise to dynamical chaos. 

\section{Conclusions}

\par
In this paper we have reviewed various definitions of quantum 
chaos. In our opinion the simplest and clearest approach is 
that of Quantum Chaology, i.e. the study of quantum systems 
which are classically chaotic. Obviously, Quantum Chaology has 
some limitations, mainly it excludes systems without a classical analog, 
but its predictions in connection to the Random Matrix Theory 
are very accurate. 
\par
Nowadays there are at least two important problems 
under investigation: 
the study of chaos without classical analog and 
the transient chaoticity of quantum systems. 
The problem of chaos in systems without a clear classical analog 
is very intricate and new ideas, like those of 
Sakata and Soloviev discussed here, are needed. 
Also the precise behaviour 
of the time scales of dynamical chaos in quantum systems 
is not fully understood but some remarks can be made. 
\par 
We observe that the limitation to persistent chaotic dynamics 
in quantum systems does not apply if the spectrum of the Hamiltonian 
operator ${\hat H}$ is continuous. In the thermodynamic limit, 
i.e. when the number $N$ of particles tends to 
infinity at constant density, the spectrum is, in general, continuous 
and true chaotic phenomena are not excluded. 
We have seen that the break time $t_B$ is very long for 
systems with many particles. 
The {\it transient chaotic dynamics} 
of quantum states and observables can be experimentally 
observed in many-body quantum systems. 
Moreover, the fact that the break time $t_B$ increases with the number 
of microscopic 
degrees of freedom explains the chaotic behaviour of macroscopic 
systems, without invoking a role for the observer or the environment. 
\par 
The study of quantum chaos for many-body or continuum 
systems (field theory) is a very promising field of research which can also 
help to better understand the foundations of quantum theory and 
statistical mechanics. 

\vskip 0.5 truecm 
\begin{center}
$*\;\;*\;\;*$
\end{center}
\vskip 0.5 truecm 
\par
One of us (L.S.) is greatly indebted to Prof. M. Robnik for many suggestions. 

\newpage 

\section*{References}

1. O. de Almeida, {\it Hamiltonian Systems: Chaos and Quantization} 
(Cambridge University Press, Cambridge, 1988).  
\\
2. M.C. Gutzwiller, {\it Chaos in Classical and Quantum 
Mechanics} (Springer-Verlag, New York, 1990). 
\\ 
3. {\it Quantum Chaos}, Ed. by H.A. Cerdeira, R. Ramaswamy, 
M.C. Gutzwiller and G. Casati (World Scientific, Singapore, 1991); 
{\it From Classical to Quantum Chaos}, Conference Proceeding SIF 
{\bf 41}, Ed. by G.F. Dell'Antonio, S. Fantoni and V.R. Manfredi 
(Editrice Compositori, Bologna, 1993). 
\\
4. {\it Quantum Chaos}, Proc. of the Int. School 
"E. Fermi", Course CXIX, Ed. by G. Casati, I. Guarneri and 
U. Smilansky (North Holland, Amsterdam, 1993). 
\\
5. {\it Quantum Chaos}, Ed. by G. Casati and B. Chirikov 
(Cambridge Univ. Press, Cambridge, 1995). 
\\
6. H. Poincar\`e, {\it Les Methodes Nouvelles de la Mechanique Celeste} 
(Gauthier-Villars, Paris, 1892). 
\\
7. A.J. Lichtenberg and M.A. Lieberman, {\it Regular and Stochastic Motion}, 
(Springer-Verlag, New York, 1983). 
\\
8. M.V. Berry, in {\it Dynamical Chaos} (The Royal Society, London, 1987). 
\\
9. M.L. Mehta, in {\it Statistical Properties of Nuclei}, Ed. by 
J.B. Garg (1972); M.L. Mehta, {\it Random Matrices}, 2nd edition  
(Academic Press, 1991). 
\\
10. M.V. Berry and M. Tabor, Proc. Roy. Soc. A {\bf 356}, 375 (1977). 
\\ 
11. M.V. Berry, Ann. Phys. {\bf 131}, 163 (1981). 
\\
12. M.V. Berry, Proc. Roy. Soc. A {\bf 400}, 229 (1985). 
\\
13. T.A. Brody, J. Flores, J.B. French, P.A. Mello, A. Pandey and 
S.S.M. Wong, Rev. Mod. Phys. {\bf 53}, 385 (1981). 
\\
14. Y. Alhassid, A. Novoselsky and N. Whelan, Phys. Rev. Lett. {\bf 65}, 
2971 (1990). 
\\
15. M.T. Lopez-Arias and V.R. Manfredi, Nuovo Cim. A {\bf 104}, 283 (1991). 
\\
16. P. Castiglione, G. Jona-Lasinio and C. Presilla, 
J. Phys. A: Math. Gen. {\bf 29}, 6169 (1996).
\\
17. M.T. Lopez-Arias, V.R. Manfredi and L. Salasnich, 
La Rivista del Nuovo Cimento {\bf 17}, n. 5 (1994); 
J.D. Garrett, J.R. German, L. Courtney and 
J.M. Espino, in {\it Future Directions in Nuclear Physics}, 
Ed. by J. Dudek and B. Hass, American Institute of Physics (1992); 
J.F. Shriner, G.E. Mitchell and T. von Egidy, 
Z. Phys. A {\bf 338}, 309 (1990). 
\\
18. E. Caurier, J.M.G. Gomez, V.R. Manfredi and L. Salasnich, Phys. Lett. B 
{\bf 365}, 7 (1996). 
\\
19. J.M.G. Gomez, V.R. Manfredi, L. Salasnich and E. Caurier, 
Phys. Rev. C {\bf 58}, 2108 (1998). 
\\
20. O. Bohigas, M.J. Giannoni and C. Schmit, 
Phys. Rev. Lett. {\bf 52}, 1 (1984). 
\\
21. M.V. Berry and M. Robnik, J. Phys. A {\bf 19}, 649 (1986). 
\\
22. M.V. Berry and M. Robnik, J. Phys. A {\bf 20}, 2389 (1987). 
\\
23. V.R. Manfredi and L. Salasnich, Z. Phys. A {\bf 343}, 1 (1992). 
\\ 
24. V.R. Manfredi, L. Salasnich and L. Dematt\`e, 
Phys. Rev. E {\bf 47}, 4556 (1993). 
\\
25. V.R. Manfredi and L. Salasnich, Int. J. Mod. Phys. E {\bf 4}, 625 (1995). 
\\
26. V.R. Manfredi, M. Rosa-Clot, L. Salasnich and S. Taddei, Int. J. Mod. 
Phys E {\bf 5}, 519 (1995). 
\\
27. L. Salasnich, Phys. Rev. D {\bf 52}, 6189 (1995). 
\\
28. L. Salasnich, Mod. Phys. Lett. A {\bf 12}, 1473 (1997). 
\\
29. L. Salasnich, Phys. Atom. Nucl. {\bf 61}, 1878 (1998). 
\\
30. R. Haq, A. Pandey and O. Bohigas, Phys. Rev. Lett. {\bf 48}, 
1086 (1982). 
\\
31. D. Delande, in {\it Chaos and Quantum Physics}, Ed. by M.J. Giannoni, 
A. Voros, J. Zinn-justin (North Holland, Amsterdam, 1991). 
\\
32. H. Ruder, G. Wunner, H. Herld and F. Geyer, 
{\it Atoms in Strong Magnetic Fields} 
(Springer-Verlag, Berlin, 1994). 
\\
33. J. Main, G. Wiebusch and K.H. Welge, in 
{\it Irregular Atomic Systems and Quantum Chaos}, 
Ed. by J.C. Gay (Gordon and Breach Science Publishers, New York, 1992). 
\\
34. T.S. Bir\`o, S.G. Matinyan and B. M\"uller, 
{\it Chaos and Gauge Field Theory} (World Scientific, Singapore, 1994). 
\\
35. F. Sakata {et al.}, Nucl. Phys. A {\bf 519}, 93c (1990). 
\\
36. P. Ring and P. Schuck, {\it The Nuclear Many-Body Problem} 
(Springer-Verlag, Berlin, 1980). 
\\
37. V.G. Soloviev, Nucl. Phys. A {\bf 554}, 77 (1993). 
\\
38. V.G. Soloviev, Nucl. Phys. A {\bf 586}, 265 (1995). 
\\
39. G. Casati, B.V. Chirikov, J. Ford and F.M. Izrailev, 
Lecture Notes in Physics. {\bf 93}, 334 (1979). 
\\
40. C. Casati, J. Ford, I. Guarneri and F. Vivaldi, 
Phys. Rev. A {\bf 34}, 1413 (1986). 
\\
41. G.P. Berman and G.M. Zaslavsky, Physica A {\bf 91}, 450 (1978). 
\\
42. A. Fetter and J. Walecka, {\it Quantum Theory 
of Many Particle Systems} (McGraw-Hill, New York, 1971). 
\\
43. G. Jona-Lasinio, C. Presilla and F. Capasso, 
Phys. Rev. Lett. {\bf 68}, 2269 (1992); G. Jona-Lasinio and C. Presilla, 
Phys. Rev. Lett. {\bf 77}, 4322 (1996). 
\\
44. M. H. Anderson {\it et al.}, Science {\bf 269}, 198 (1995); 
C. C. Bradley {\it et al.}, Phys. Rev. Lett. {\bf 75}, 1687 (1995); 
K. B. Davis {\it et al.}, Phys. Rev. Lett. {\bf 75}, 3969 (1995). 
\\
45. L. Reatto, A. Parola and L. Salasnich, 
J. Low Temp. Phys. {\bf 113}, 195 (1998). 
\\
46. E. Cerboneschi, R. Mannella, E. Arimondo and L. Salasnich, 
Phys. Lett. A {\bf 249}, 245 (1998). 

\newpage

\section*{Figure Captions}

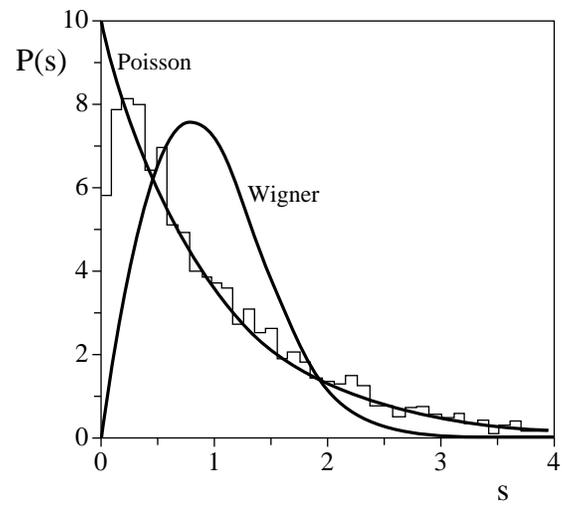
{\bf Figure 1}: $P(s)$ for "cold" deformed rare-earth nuclei 
(adapted from Ref. 17). 
\\
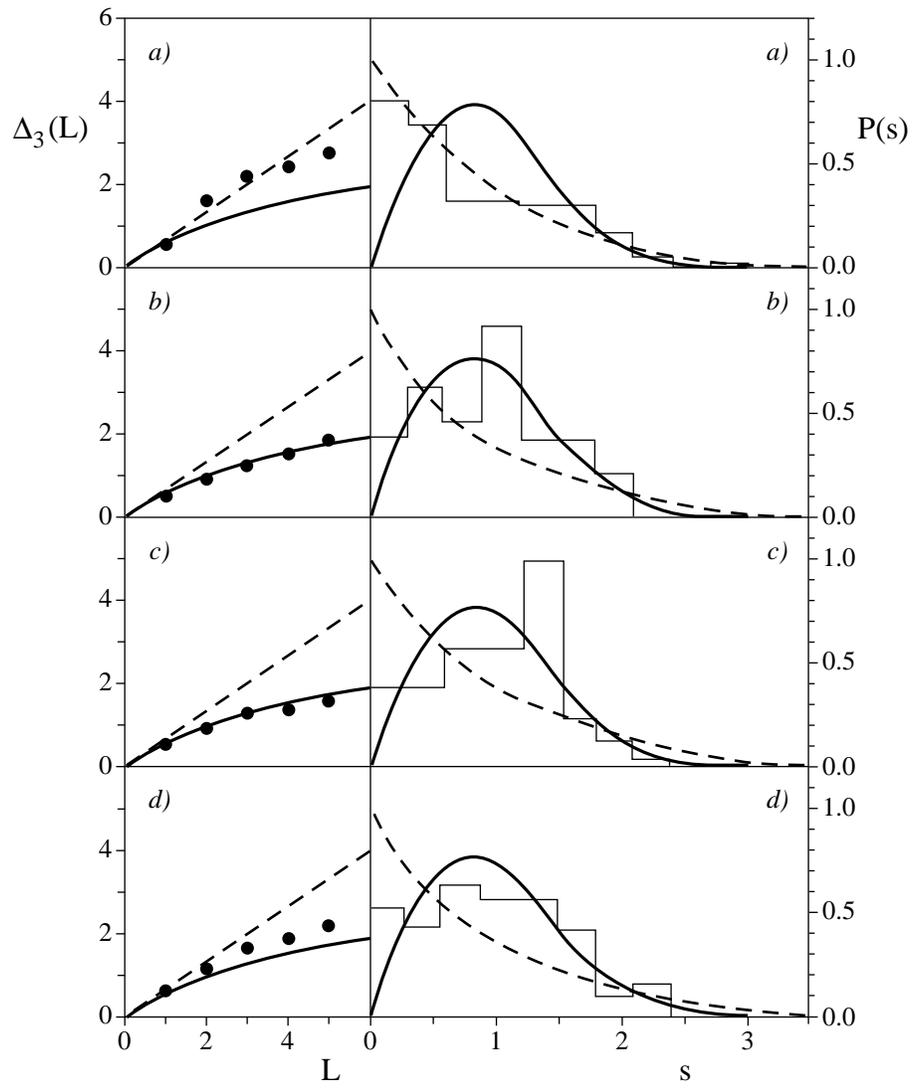
{\bf Figure 2}: Spectral statistics for nuclei with atomic mass 
$24<A<244$ and excitation energy of few MeV. 
a) $2+$ and $4^+$ states, even-even nuclei; 
b) all other states, even-even nuclei; 
c) states with non-natural parity, odd-odd nuclei; 
d) states with natural parity, odd nuclei 
(adapted from Ref. 17). 
\\
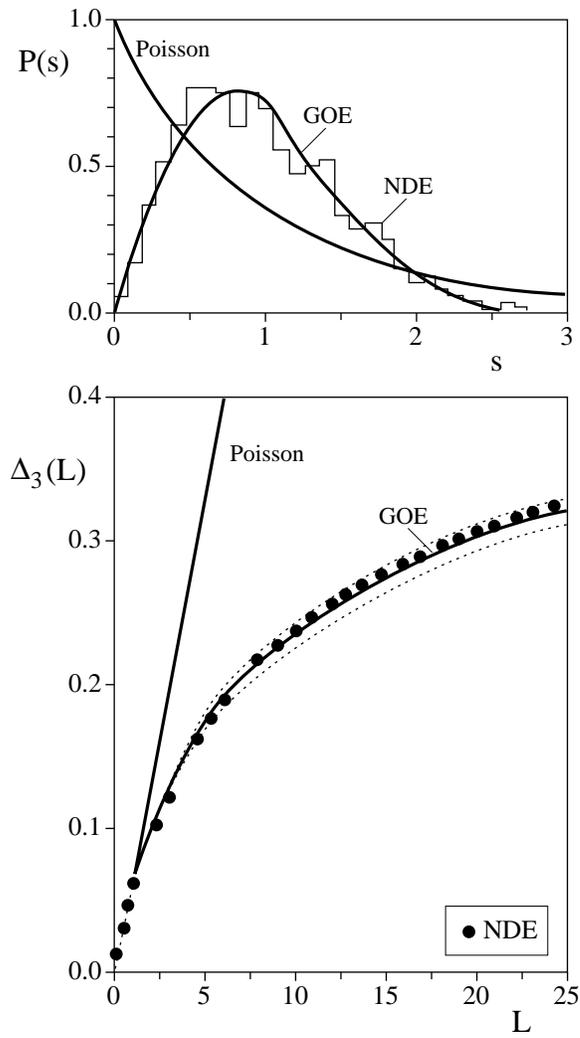
{\bf Figure 3}: Comparison of the nearest-neighbour spacing distribution 
$P(s)$ and spectral rigidity $\Delta_3(L)$ of the Nuclear 
Data Ensemble (NDE) with the GOE predictions 
(adapted from Ref. 30). 
\\
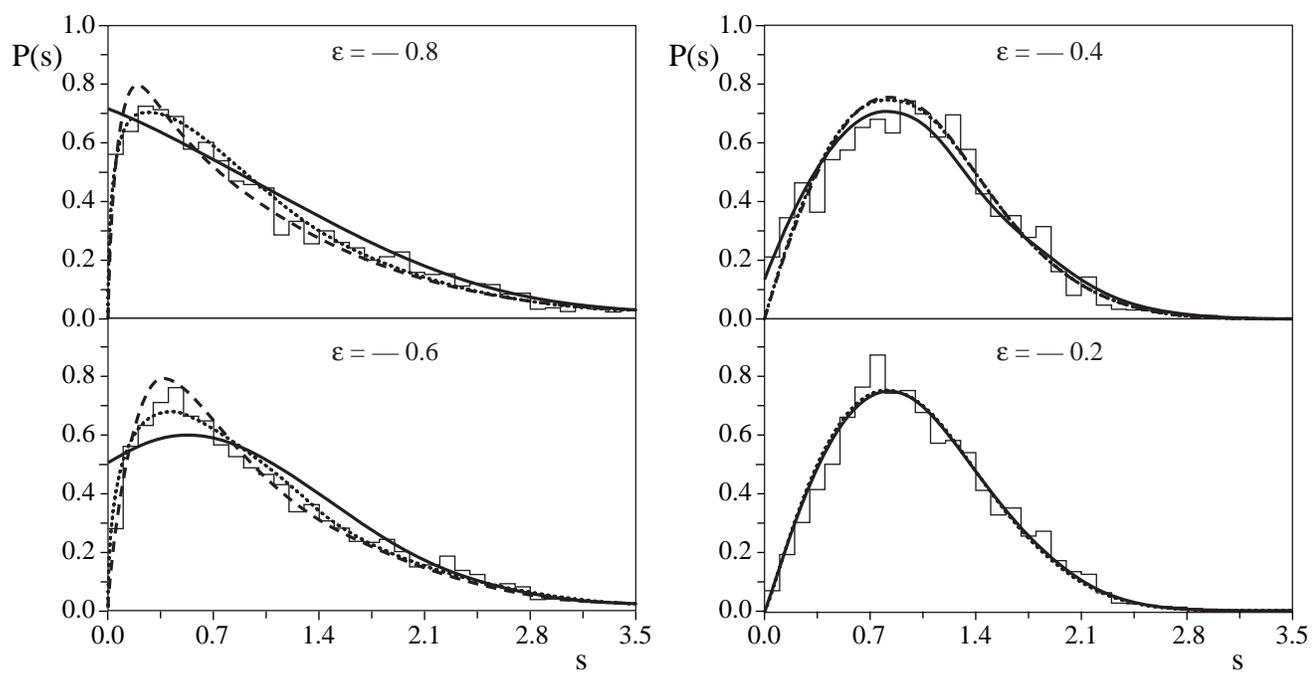
{\bf Figure 4}: Histograms of the level distances of 
quantum energies in the Hydrogen 
atom in magnetic fields at different values of the scaled energy. 
The smooth curves are the results of the fits to the histograms 
(adapted from Ref. 32). 
\\
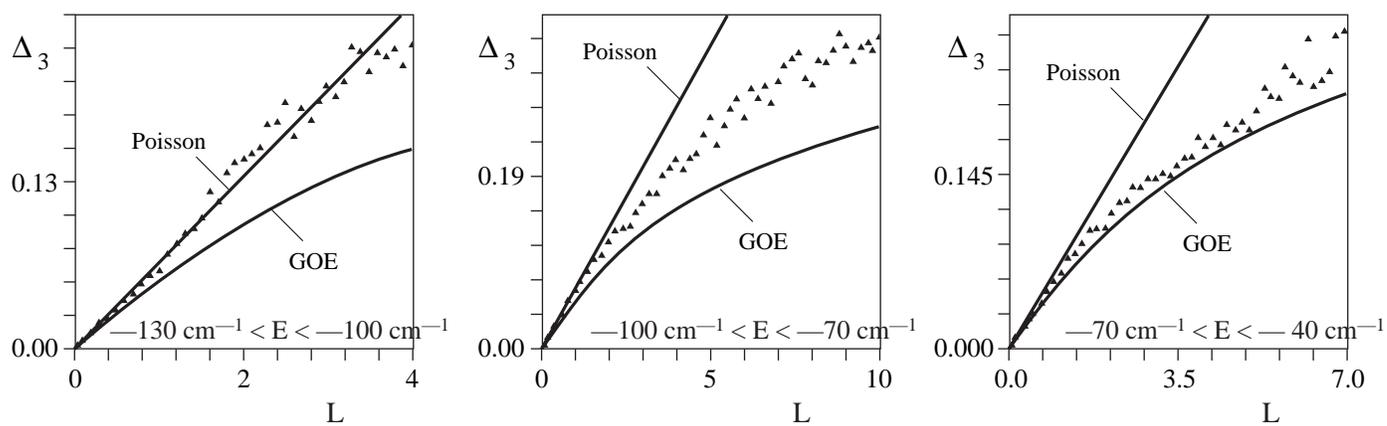
{\bf Figure 5}: Spectral rigidity $\Delta_3(L)$ for energy level 
sequences of the Hydrogen atom in a magnetic field of $6$ Tesla 
in three different energy intervals. 
The transition to the GOE distribution as soon as the 
classical motion becomes chaotic is also visible here 
(adapted from Ref. 32). 
\\
{\bf Figure 6}: The Poincar\`e sections of the model. From the top: 
$v=1$, $v=1.1$ and $v=1.2$. Energy $E = 10$ and interaction $g=1$ 
(adapted from Ref. 28). 
\\
{\bf Figure 7}: $P(s)$ distribution. From the top: 
$v=1$ ($\omega=0.92$), $v=1.1$ ($\omega =0.34$) and $v=1.2$ ($\omega =0.01$), 
where $\omega$ is the Brody parameter. First 100 energy levels 
and interaction $g=1$. The dotted, dashed and solid curves stand 
for Wigner, Poisson and Brody distributions, respectively 
(adapted from Ref. 28). 
\\
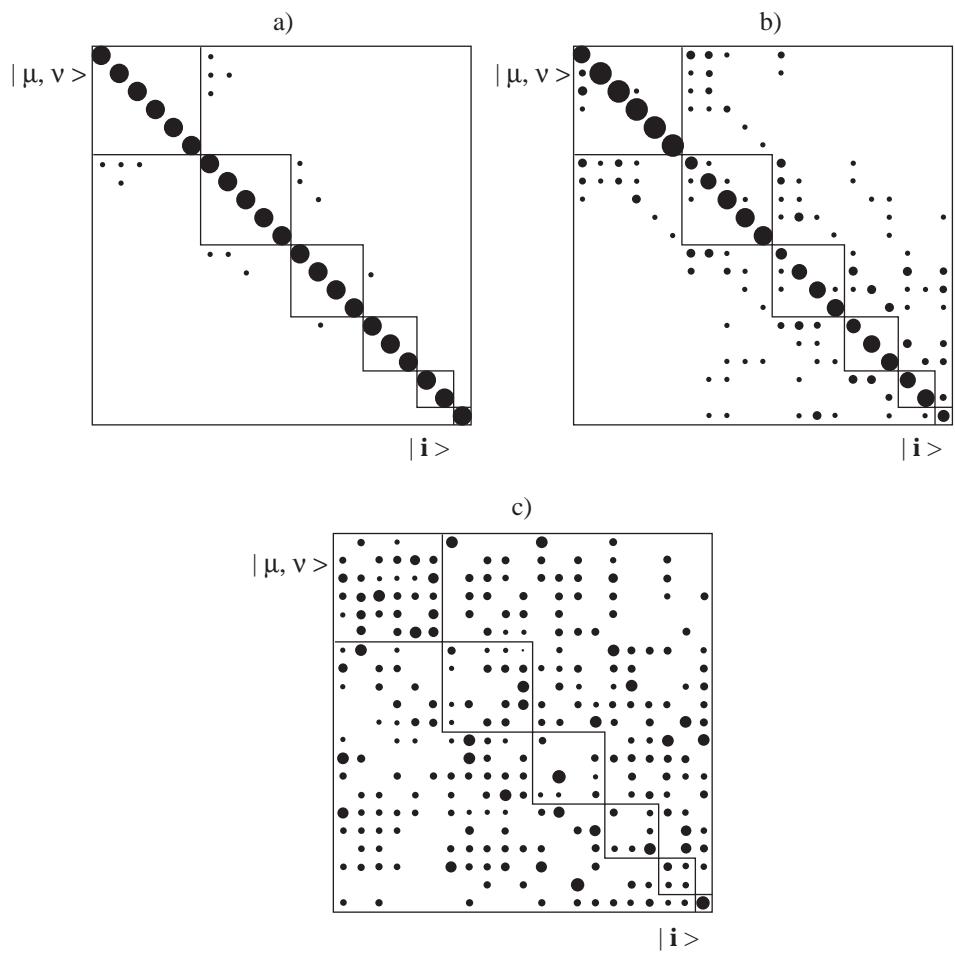
{\bf Figure 8}: Exact eigenstates $|i>$ expressed in the $|\mu , \nu>$ 
basis states: 
a) Quantum Integrable States;  
b) Quantum KAM States; 
c) Quantum Chaotic States (adapted from Ref. 35). 
\\
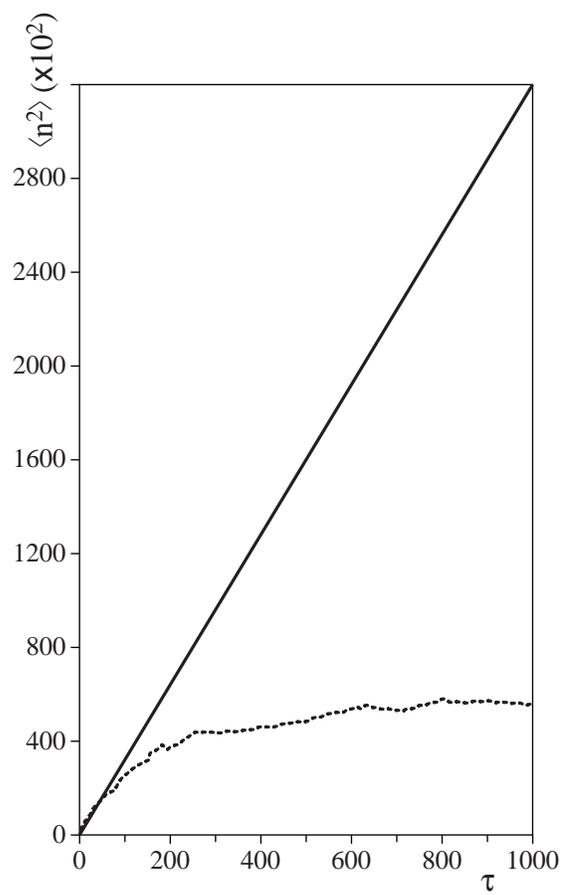
{\bf Figure 9}: Classical (solid curve) and quantum (dotted curve) 
unperturbed energy $<n^2(\tau)>=2E$ 
as a function of time $\tau$ (number of map iterations) 
(adapted from Ref. 5). 

\end{document}